\documentclass[epj]{svjour} 
\usepackage{epsf,cite,epsfig,psfig,float,amssymb,stmaryrd,latexsym} 
\usepackage{graphics,psfrag}
\usepackage{graphicx} 
\newcommand{\be}{\begin{equation}} 
\newcommand{\ee}{\end{equation}} 
\newcommand{\beqa}{\begin{eqnarray}} 
\newcommand{\eeqa}{\end{eqnarray}}

\begin{document} 
\titlerunning{On the chiral effective meson-baryon Lagrangian at third order}
\title{On the chiral effective meson-baryon Lagrangian at third order
\thanks{This
research is part of the EU Integrated Infrastructure Initiative Hadron 
Physics Project under contract number RII3-CT-2004-506078. 
Work supported in part by DFG (SFB/TR 16 ``Subnuclear Structure of Matter'').
}} 
 
\author{Matthias Frink \inst{1} \thanks{Electronic address:~mfrink@itkp.uni-bonn.de}  
\and 
Ulf-G. Mei{\ss}ner \inst{1}\inst{2} \thanks{ 
Electronic address:~meissner@itkp.uni-bonn.de}
}                     
%
%
\institute{  
Helmholtz-Institut f\"ur Strahlen- und Kernphysik (Theorie), 
Universit\"at Bonn, Nu{\ss}allee 14-16, D-53115 Bonn, Germany 
\and 
Forschungszentrum J{\" u}lich, Institut f{\" u}r Kernphysik 
(Theorie),  D-52425  J{\" u}lich, Germany 
} 
\date{Received: date / Revised version: date} 
%
\abstract{ 
We show that the recently constructed complete and 
``minimal'' third order meson--baryon effective chiral
Lagrangian can be further reduced from 84 to 78 independent operators.
} 
\maketitle 
%


Recently, Oller et al. \cite{Oller:2006yh} 
have presented a complete and ``minimal''
chiral effective Lagrangian for three flavor baryon chiral perturbation
theory to third order in the chiral expansion, utilizing the methods 
outlined in Ref.~\cite{Fettes:2000gb}
for constructing the fourth order two flavor effective 
Lagrangian. The first attempt to construct the SU(3) meson-baryon Lagrangian
goes back to Krause \cite{Krause:1990xc}.
However, he made no attempt at minimizing the
number of independent operators at third order. We are presently constructing
the minimal fourth order SU(3) Lagrangian for Goldstone bosons coupled to the
ground state baryon octet and external sources. As a by--product we show in
this note that the number of 84 independent third order operators 
in Ref.~\cite{Oller:2006yh}
can further be reduced by a combination of using certain
Cayley--Hamilton relations and the baryon equations of motion.

The chiral effective Lagrangian can be constructed by writing down the
pertinent building blocks to generate chirally invariant operators at a given
order in the chiral expansion. The building blocks are expressed in terms 
of the standard meson matrices $U,~u^2=U$, the baryon matrix $B$, and the 
external  sources $\chi=s+ip,~r_\mu,~l_\mu$:
\begin{eqnarray}
u_\mu&=& i u^\dagger \nabla_\mu U u^\dagger=-i u \nabla_\mu U^\dagger u~,\nonumber\\
\chi^\pm&=& u \chi^\dagger u \pm u^\dagger \chi u^\dagger~,\nonumber\\
F^\pm_{\mu \nu}&=& u^\dagger R_{\mu \nu} u \pm u L_{\mu \nu} u^\dagger~.
\end{eqnarray}
Here
\begin{eqnarray}
\nabla_\mu U &=&\partial_\mu U-i r_\mu U+i U l_\mu~,\nonumber\\
R_{\mu\nu}&=&\partial_\mu r_\nu-\partial_\nu r_\mu-i\left[r_\mu,r_\nu\right]~,
\end{eqnarray}
in terms of the commutator $\left[~~,~\right]$, and similarly for $l_\mu$.
The covariant derivative acting on any field $X=B,u_\mu,\ldots$ reads
\begin{equation}
\left[D_\mu,X\right]=\partial_\mu X +\left[\Gamma_\mu,X\right]~,
\end{equation}
where $\Gamma_\mu$ is given by 
\begin{equation}
\Gamma_\mu=\frac{1}{2} \left[u^\dagger,\partial_\mu u\right]
-\frac{i}{2}\left(u^\dagger r_\mu u+u l_\mu u^\dagger\right)~.
\end{equation}
To proceed, one has to assign a chiral dimension to these building blocks.
We work here in standard chiral perturbation theory, that is all fields
are of order $q^0$ (where $q$ is the small expansion parameter), and so are derivatives acting on baryon fields. 
Derivatives applied to meson or external fields, vector and axial--vector sources are
${\cal O}(q)$, and the field strength tensors as well as the scalar and
pseudoscalar sources are ${\cal O}(q^2)$. The latter assignment reflects the
fact that the Goldstone boson masses squared are proportional to the quark
masses at leading order in the chiral expansion. Consequently, the leading
order effective Lagrangian is of chiral order one, next-to-leading order
corrections are of order two, and the first loop contributions appear at
third order together with the local contact interactions that are considered
in this note. Thus, the effective Lagrangian takes the form
\begin{equation}
{\cal L}_{MB} = {\cal L}_{MB}^{(1)} +  {\cal L}_{MB}^{(2)} + {\cal
  L}_{MB}^{(3)} + \ldots~,
\end{equation} 
where the ellipsis denotes higher order terms.

We skip here any technical details how to construct the effective Lagrangian
from the building blocks, apart from certain relations that are utilized to
further minimize the number of third order operators. For these details,
we refer to Refs.~\cite{Fettes:2000gb,Oller:2006yh,FM4}.
We fully agree with ${\cal L}^{(2)}_{M B}$ in Eq.~(5.1) of
Ref.~\cite{Oller:2006yh}. 
In ${\cal L}^{(3)}_{M B}$ in Eq.~(5.2), however, six of the 84 terms 
listed there are redundant by one of the following mechanisms:
\begin{itemize}
\item SU(3) Cayley-Hamilton trace identities;
\item the lowest-order baryon equation of motion\\ 
$i\gamma^\mu \left[D_\mu,B\right]=m_0 B+{\cal O}(q)$, with $m_0$ the
octet baryon mass in the chiral limit. 
\end{itemize}
The Cayley-Hamilton identities allow to discard five of the structures in 
${\cal L}^{(3)}_{M B}$ in Ref.~\cite{Oller:2006yh}, 
e.g.~$O_{20}$, $O_{21}$, $O_{22}$, $O_{35}$, $O_{36}$.
The  details of this procedure are spelled out in App.~\ref{app:CH}.

We now give details how a further reduction of the number of terms is achieved.
By means of the lowest-order baryon equation of motion monomials containing 
$\gamma^\mu \left[D_\mu,B\right]$ can be reduced to the corresponding
structures without the covariant derivative and higher-order contributions. 
Revealing prospective Lagrangian terms to be redundant means exposing those 
$\gamma^\mu \left[D_\mu,B\right]$ structures where-ever they occur hidden in 
the Clifford-algebra. In Ref.~\cite{Oller:2006yh} this procedure is implemented in 
terms of the relations Eqs.~(4.4)--(4.10). Some of these equations are,
however, interrelated and their full content cannot be unraveled by treating 
them as separate identities. To illustrate this feature consider the following 
third-order combination of building blocks comprising two mesonic fields 
$u_\lambda,u_\rho$ together with two covariant derivatives, where one 
derivative $\tilde{D}_\nu$ acts on meson fields, as indicated by the 
tilde symbol, and the other one, $D_\mu$, acts on the baryon matrix:
\begin{equation}\label{eq:uu}
(u_\lambda,u_\rho,\tilde{D}_\nu,D_\mu)\leftrightarrow 
(\varepsilon^{\mu\nu\rho\lambda}, g^{\mu \nu} \sigma^{\rho \lambda},
g^{\mu \nu} g^{\rho \lambda},\ldots)
\end{equation}
The left-hand-side gives rise to invariants by combing meson and baryon
matrices into products of traces and applying the covariant derivatives. 
Having one derivative each act on the two types of fields involved, one
specific trace configuration generates contributions referred to as 
$D_\mu \tilde{D}_\nu$-terms and accordingly for $\tilde{D}_\mu D_\nu$.
As suggested by the right-hand-side of Eq.~(\ref{eq:uu}), 
the four Lorentz indices can be contracted with the totally 
anti-symmetric tensor $\varepsilon$, a product of one metric tensor $g$ and 
one anti-symmetric Clifford-algebra element $\sigma$, or else two metric 
tensors, where the ellipsis indicates that any ordering of the indices is 
possible, up to occurring index symmetries. Consider now the two structures 
$\sigma^{\mu \rho} g^{\nu \lambda}$, $\, g^{\mu \rho} \sigma^{\nu \lambda}$. 
The following identities among Clifford-algebra elements,
\begin{eqnarray}
g^{\mu \rho} g^{\nu \lambda}&=&-i\sigma^{\mu \rho} g^{\nu \lambda}
+\gamma^\rho \gamma^\mu g^{\nu \lambda}\nonumber\\
&=&-ig^{\mu \rho} \sigma^{\nu \lambda}+ g^{\mu \rho}\gamma^\lambda \gamma^\nu~,
\end{eqnarray}
are the basis of Eq.~(4.4) in Ref.~\cite{Oller:2006yh}, which implies that 
$\sigma^{\mu \rho} g^{\nu \lambda}$ in connection with the $D_\mu
\tilde{D}_\nu$-terms 
is eliminated in favor of $g^{\mu \rho} g^{\nu \lambda} $, and accordingly 
for $ g^{\mu \rho} \sigma^{\nu \lambda}$ contracted with the 
$\tilde{D}_\mu D_\nu$-configurations. If all $D_\mu \tilde{D}_\nu$- 
as well as $\tilde{D}_\mu D_\nu$-terms are summed up, the covariant
derivatives can be reshuffled by a total derivative argument to act on baryon 
fields only, thus producing a $D_\mu D_\nu$-structure.  
$\sigma^{\mu \rho} g^{\nu \lambda}$ contracted with this structure is 
related by the eliminating index $\mu$ to $g^{\mu \rho} g^{\nu \lambda} $, 
which, in turn, can be connected to the corresponding contribution 
including $g^{\mu \rho} \sigma^{\nu \lambda}$ by means of manipulations in 
the index $\nu$. Therefore, the above $D_\mu D_\nu$-term contracted with the 
difference $\sigma^{\mu \rho} g^{\nu \lambda}- g^{\mu \rho} \sigma^{\nu
  \lambda}$ 
can be ignored making use of the baryon equations of motion. 
As $D_\mu D_\nu$ is symmetric in the indices $\mu$ and $\nu$, this leads 
to the elimination of one $D_\mu \tilde{D}_\nu$- or 
$\tilde{D}_\mu D_\nu$-contribution associated with a field configuration 
anti-symmetric in the indices $\rho$ and $\lambda$ contracted with, 
say, $ \sigma^{\mu \rho} g^{\nu \lambda}$. Since this Clifford structure 
reduces the combination $ \left[D_\mu,u_\rho\right]$ via the 
relation $\left[D_\mu,u_\rho\right]-\left[D_\rho,u_\mu\right]=F_{\mu \rho}^-$, 
we are left with two prospective terms including $
\left[D_\mu,u_\lambda\right]$ which are not independent: 
one of the monomials $O_9$, $O_{10}$ in ${\cal L}^{(3)}_{M B}$ 
from Ref.~\cite{Oller:2006yh} is redundant. Thus, together with the five 
operators removed by use of the Cayley-Hamilton relations, we have reduced the
number of independent operators from 84 to 78.

We furthermore remark that $O_{41}$ in Ref.~\cite{Oller:2006yh} has the wrong 
behavior under charge conjugation. Adopting the notation used there the 
operation of charge conjugation is given by:
\begin{eqnarray}
&&\langle\overline B\sigma^{\lambda \tau} D_\rho B\rangle\langle u^\mu u^\nu u^\rho\rangle \varepsilon_{\mu \nu \lambda\tau}\nonumber\\&&\quad\quad\stackrel{\cal C}{\longrightarrow}-\langle\overline B\stackrel{ \leftarrow}{D}_\rho\sigma^{\lambda \tau} B\rangle\langle u^\nu u^\mu u^\rho\rangle \varepsilon_{\mu \nu  \lambda\tau}~,
\end{eqnarray}
where use has been made of the cyclic property of the trace. Accounting for 
the inverted order of fields in the final trace by an index exchange induces 
a change in the relative sign. The proper charge conjugation invariant 
structure is of higher order and the remark stating an abnormal 
non-relativistic power counting behavior for this term is then rendered pointless.

Similarly, the ordering of indices in the monomials $O_{31}$, $O_{33}$, and
$O_{34}$ does not match the conditions imposed by charge conjugation. 
In the notation used there, the invariant assignment of indices e.g.~for 
$ O_{31}$ reads, given the symmetric product of covariant derivatives:
\begin{equation}
\langle\overline B \gamma_5 \gamma_\mu D_{\nu\rho}B u^\mu u^\nu u^\rho\rangle
+\langle\overline B \stackrel{\leftarrow}{D}_{\nu\rho} \gamma_5 
\gamma_\mu B u^\nu u^\rho u^\mu\rangle~.
\end{equation}

We finally present a list of the Lagrangian invariants where the above 
corrections have been accounted for:
\begin{equation}
{\cal L}^{(3)}_{M B}=\sum_{i=1}^{78} \, d_i \, O^{(3)}_i~.
\end{equation}
The explicit terms are collected in Tab.~\ref{table:1}.
The basis chosen is different from the one in Ref.~\cite{Oller:2006yh}. 
All third-order LECs $d_i$ accompanying the operators $O_i^{(3)}$
are of dimension mass$^{-2}\,$. Using the same techniques, we
are presently constructing the complete and minimal fourth order
La{\-}grang{\-}ian for SU(3) baryon chiral perturbation theory. The terms
are needed for any complete one--loop calculation (for the
example of the baryon masses at fourth order, see e.g.~Ref.~\cite{Frink:2004ic}
and references therein).


\section*{Acknowledgements}
We are grateful to Bastian Kubis for useful comments.

\appendix 
\def\theequation{\Alph{section}.\arabic{equation}} 
\setcounter{equation}{0} 

\section{Cayley-Hamilton relations}
\label{app:CH}

 The following notation is introduced to classify Cayley-Hamilton identities: 
a relation consisting of a vanishing sum of traces and  products thereof 
is referred to by the term in the sum that contains the trace
extending  over the maximum number of matrices. Applied to the case of 
three matrix-valued fields $A_1,A_2,A_3$ included, together with the baryon 
fields $\overline B,B$, in one or several traces, 15 independent identities 
result for complex $3\times3$ matrices. Formulated as invariants with definite 
behavior under charge conjugation, they fall into C-even and C-odd relations, 
respectively. These are listed below together with a possible choice of
structures thus rendered redundant. Here, $A_1 A_2 A_3 + {\rm perm.}$ stands 
for the product of the three matrices in all possible orders given by the 
six possible permutations of the indices $\left\{1,2,3\right\}$ summed up. 
$\left[~~,~\right]$ and $\left\{~~,~\right\}$ stand for commutators and 
anti-commutators, respectively. $\Gamma$ represents an element of the Clifford-algebra.

\medskip

\noindent
C-even:

\begin{equation}
{\rm Tr}\left\{(\overline B B A_1+{\rm perm.})A_2\right\} {\rm
  Tr}\left\{A_3\right\}
\end{equation}
is used to eliminate
\begin{eqnarray} &&{\rm Tr}\left\{\overline B A_1\right\} \Gamma {\rm Tr}\left\{A_2 B\right\}{\rm Tr}\left\{A_3\right\}\nonumber\\
&&\quad + {\rm Tr}\left\{\overline B  A_2\right\} \Gamma {\rm Tr}\left\{A_1
  B\right\}{\rm Tr}\left\{A_3\right\}~,
\end{eqnarray}
and accordingly for ${\rm Tr}\left\{(\overline B B A_1+{\rm perm.})A_3\right\}
{\rm Tr}\left\{A_2\right\}$ and ${\rm Tr}\left\{(\overline B B A_2+{\rm
    perm.})A_3\right\} {\rm Tr}\left\{A_1\right\}\,$.

\begin{eqnarray}
&& {\rm Tr}\left\{(\overline B B A_1+{\rm perm.})\left\{A_2,A_3\right\}\right\}~,\nonumber\\
&& {\rm Tr}\left\{(\overline B B A_2+{\rm perm.})\left\{A_1,A_3\right\}\right\}~,\nonumber\\
&& {\rm Tr}\left\{(\overline B B A_3+{\rm perm.})\left\{A_1,A_2\right\}\right\}~,\nonumber\\
&& {\rm Tr}\left\{( A_1 A_2 A_3+{\rm perm.})\left[\overline B,B\right]\right\}~,\nonumber\\
&& {\rm Tr}\left\{( B A_2 A_3+{\rm perm.})\left[\overline B,A_1\right]\right\}\nonumber\\
&&\quad+{\rm Tr}\left\{(\overline B A_2 A_3+{\rm perm.})\left[A_1,B\right]\right\}~,\nonumber\\&& {\rm Tr}\left\{( B A_1 A_3+{\rm perm.})\left[\overline B,A_2\right]\right\}\nonumber\\
&&\quad+{\rm Tr}\left\{(\overline B A_1 A_3+{\rm perm.})\left[A_2,B\right]\right\}~,\nonumber\\&& {\rm Tr}\left\{( B A_1 A_2+{\rm perm.})\left\{\overline B,A_3\right\}\right\}\nonumber\\
&&\quad+{\rm Tr}\left\{(\overline B A_1 A_2+{\rm
 perm.})\left\{A_3,B\right\}\right\}
\end{eqnarray}

is used to eliminate

\begin{eqnarray}
&&{\rm Tr}\left\{\overline B \Gamma  \left\{A_1,B\right\}\right\} {\rm Tr}\left\{A_2 A_3\right\}~,\nonumber\\
&& {\rm Tr}\left\{\overline B \Gamma  \left\{A_2,B\right\}\right\} {\rm Tr}\left\{A_1 A_3\right\}~,\nonumber\\
&& {\rm Tr}\left\{\overline B \Gamma  \left\{A_3,B\right\}\right\} {\rm Tr}\left\{A_1 A_2\right\}~,\nonumber\\&&{\rm Tr}\left\{\overline B \Gamma  \left[A_1,B\right]\right\} {\rm Tr}\left\{A_2 A_3\right\}~,\nonumber\\
&& {\rm Tr}\left\{\overline B \Gamma  \left[A_2,B\right]\right\} {\rm Tr}\left\{A_1 A_3\right\}~,\nonumber\\
&& {\rm Tr}\left\{\overline B \Gamma  \left[A_3,B\right]\right\} {\rm Tr}\left\{A_1 A_2\right\}~,\nonumber\\
&& {\rm Tr}\left\{\overline B A_3\right\} \Gamma {\rm Tr}\left\{\left\{A_1,A_2\right\} B\right\}\nonumber\\
&&\quad+ {\rm Tr}\left\{\overline B\left\{A_1,A_2\right\}  \right\} \Gamma {\rm Tr}\left\{A_3 B\right\}~,
\end{eqnarray}
where the final relation is understood in its totally symmetric combination.
\\
C-odd:\begin{eqnarray}
&& {\rm Tr}\left\{(\overline B B A_1+{\rm perm.})\left[A_2,A_3\right]\right\}~,\nonumber\\
&& {\rm Tr}\left\{(\overline B B A_2+{\rm perm.})\left[A_1,A_3\right]\right\}~,\nonumber\\
&& {\rm Tr}\left\{(\overline B B A_3+{\rm perm.})\left[A_1,A_2\right]\right\}~,\nonumber\\
&& {\rm Tr}\left\{( B A_2 A_3+{\rm perm.})\left\{\overline B,A_1\right\}\right\}\nonumber\\
&&\quad-{\rm Tr}\left\{(\overline B A_2 A_3+{\rm perm.})\left\{A_1,B\right\}\right\}~,\nonumber\\&& {\rm Tr}\left\{( B A_1 A_3+{\rm perm.})\left\{\overline B,A_2\right\}\right\}\nonumber\\
&&\quad-{\rm Tr}\left\{(\overline B A_1 A_3+{\rm
    perm.})\left\{A_2,B\right\}\right\}
\end{eqnarray}

is used to eliminate

\begin{eqnarray}
&& {\rm Tr}\left\{\overline B A_1\right\} \Gamma {\rm Tr}\left\{\left[A_2,A_3\right] B\right\}\nonumber\\
&&\quad+ {\rm Tr}\left\{\overline B\left[A_2,A_3\right]  \right\} \Gamma {\rm Tr}\left\{A_1 B\right\}~,\nonumber\\
&& {\rm Tr}\left\{\overline B A_2\right\} \Gamma {\rm Tr}\left\{\left[A_1,A_3\right] B\right\}\nonumber\\
&&\quad+ {\rm Tr}\left\{\overline B\left[ A_1,A_3\right]   \right\} \Gamma {\rm Tr}\left\{A_2 B\right\}~,\nonumber\\
&& {\rm Tr}\left\{\overline B A_3\right\} \Gamma {\rm Tr}\left\{\left[ A_1,A_2\right]  B\right\}\nonumber\\
&&\quad+ {\rm Tr}\left\{\overline B\left[ A_1,A_2\right]   \right\} \Gamma {\rm Tr}\left\{A_3 B\right\}~,\nonumber\\
&& {\rm Tr}\left\{\overline B A_1\right\} \Gamma {\rm Tr}\left\{\left\{A_2,A_3\right\}B\right\}\nonumber\\
&&\quad- {\rm Tr}\left\{\overline B\left\{A_2,A_3\right\} \right\} \Gamma {\rm Tr}\left\{A_1 B\right\}~,\nonumber\\
&& {\rm Tr}\left\{\overline B A_2\right\} \Gamma {\rm Tr}\left\{\left\{A_1,A_3\right\} B\right\}\nonumber\\
&&\quad- {\rm Tr}\left\{\overline B\left\{ A_1,A_3\right\}   \right\} \Gamma {\rm Tr}\left\{A_2 B\right\}~,
\end{eqnarray}
where the final two relations are understood in their $ 1\leftrightarrow 3$ anti-symmetric and $ 2\leftrightarrow 3$ anti-symmetric combinations, respectively.

\appendix 
\def\theequation{\Alph{section}.\arabic{equation}} 
\setcounter{equation}{0}

\begin{table*}[htp]
\begin{center}
\begin{tabular}{|r|l|}\hline 
$i$&$O^{(3)}_i$\\ \hline\hline
&\\
$1$&$m_0 {\rm Tr}\left\{\overline B \gamma^5 \left[\chi^-,B\right]\right\}$\\
$2$&$m_0 {\rm Tr}\left\{\overline B \gamma^5 \left\{\chi^-,B\right\}\right\}$\\
$3$&$m_0 {\rm Tr}\left\{\overline B \gamma^5 B\right\} {\rm Tr}\left\{\chi^-\right\}$\\
$4$&$ i\Big({\rm Tr}\left\{\overline B \varepsilon^{\mu\nu\rho\tau}\gamma_\tau \left[\left[u_\mu,u_\nu\right],\left[u_\rho, B\right]\right]\right\}+ {\rm Tr}\left\{\overline B \varepsilon^{\mu\nu\rho\tau}\gamma_\tau  \left[u_\rho,\left[\left[u_\mu,u_\nu\right], B\right]\right]\right\}\Big)$\\
$5$&$ i\Big({\rm Tr}\left\{\overline B \varepsilon^{\mu\nu\rho\tau}\gamma_\tau \left[\left[u_\mu,u_\nu\right],\left\{u_\rho, B\right\}\right]\right\}+ {\rm Tr}\left\{\overline B \varepsilon^{\mu\nu\rho\tau}\gamma_\tau  \left\{u_\rho,\left[\left[u_\mu,u_\nu\right], B\right]\right\}\right\}\Big)$\\
$6$&$ i\Big({\rm Tr}\left\{\overline B \varepsilon^{\mu\nu\rho\tau}\gamma_\tau \left\{\left[u_\mu,u_\nu\right],\left[u_\rho, B\right]\right\}\right\}+ {\rm Tr}\left\{\overline B \varepsilon^{\mu\nu\rho\tau}\gamma_\tau  \left[u_\rho,\left\{\left[u_\mu,u_\nu\right], B\right\}\right]\right\}\Big)$\\
$7$&$ i\Big({\rm Tr}\left\{\overline B \varepsilon^{\mu\nu\rho\tau}\gamma_\tau \left\{\left[u_\mu,u_\nu\right],\left\{u_\rho, B\right\}\right\}\right\}+ {\rm Tr}\left\{\overline B \varepsilon^{\mu\nu\rho\tau}\gamma_\tau  \left\{u_\rho,\left\{\left[u_\mu,u_\nu\right], B\right\}\right\}\right\}\Big)$\\
$8$&$ i {\rm Tr}\left\{\overline B \varepsilon^{\mu\nu\rho\tau}\gamma_\tau B\right\} {\rm Tr}\left\{u_\mu\left[u_\nu,u_\rho\right]\right\}$\\
$9$&${\rm Tr}\left\{\overline B \gamma^\mu\gamma^5 \left[u_\nu,\left[u^\nu,\left[u_\mu, B\right]\right]\right]\right\}+{\rm Tr}\left\{\overline B \gamma^\mu\gamma^5 \left[u_\mu,\left[u_\nu,\left[u^\nu, B\right]\right]\right]\right\}$\\ 
$10$&${\rm Tr}\left\{\overline B \gamma^\mu\gamma^5 \left[u_\nu,\left[u^\nu,\left\{u_\mu, B\right\}\right]\right]\right\}+{\rm Tr}\left\{\overline B \gamma^\mu\gamma^5 \left\{u_\mu,\left[u_\nu,\left[u^\nu, B\right]\right]\right\}\right\}$\\ 
$11$&${\rm Tr}\left\{\overline B \gamma^\mu\gamma^5 \left[u_\nu,\left\{u^\nu,\left[u_\mu, B\right]\right\}\right]\right\}+{\rm Tr}\left\{\overline B \gamma^\mu\gamma^5 \left[u_\mu,\left\{u_\nu,\left[u^\nu, B\right]\right\}\right]\right\}$\\ 
$12$&${\rm Tr}\left\{\overline B \gamma^\mu\gamma^5 \left[u_\nu,\left\{u^\nu,\left\{u_\mu, B\right\}\right\}\right]\right\}+{\rm Tr}\left\{\overline B \gamma^\mu\gamma^5 \left\{u_\mu,\left\{u_\nu,\left[u^\nu, B\right]\right\}\right\}\right\}$\\ 
$13$&${\rm Tr}\left\{\overline B \gamma^\mu\gamma^5 \left\{u_\nu,\left\{u^\nu,\left[u_\mu, B\right]\right\}\right\}\right\}+{\rm Tr}\left\{\overline B \gamma^\mu\gamma^5 \left[u_\mu,\left\{u_\nu,\left\{u^\nu, B\right\}\right\}\right]\right\}$\\ 
$14$&${\rm Tr}\left\{\overline B \gamma^\mu\gamma^5 \left\{u_\nu,\left\{u^\nu,\left\{u_\mu, B\right\}\right\}\right\}\right\}+{\rm Tr}\left\{\overline B \gamma^\mu\gamma^5 \left\{u_\mu,\left\{u_\nu,\left\{u^\nu, B\right\}\right\}\right\}\right\}$\\
$15$&${\rm Tr}\left\{\overline B \gamma^\mu\gamma^5 \left[ \left[u_\nu,\left[u^\nu,u_\mu\right]\right], B\right]\right\}$\\
$16$&${\rm Tr}\left\{\overline B \gamma^\mu\gamma^5 \left\{ \left[u_\nu,\left[u^\nu,u_\mu\right]\right], B\right\}\right\}$\\
$17  $&${\rm Tr}\left\{\overline B u_\nu\right\} \gamma^\mu\gamma^5 {\rm Tr}\left\{\left\{u^\nu,u_\mu\right\} B\right\}+ {\rm Tr}\left\{\overline B\left\{u_\nu,u_\mu\right\}  \right\} \gamma^\mu\gamma^5 {\rm Tr}\left\{u^\nu B\right\}$\\
$18$&${\rm Tr}\left\{\overline B u_\nu\right\} \gamma^\mu\gamma^5 {\rm Tr}\left\{\left[u^\nu,u_\mu\right] B\right\}- {\rm Tr}\left\{\overline B\left[u_\nu,u_\mu\right]  \right\} \gamma^\mu\gamma^5 {\rm Tr}\left\{u^\nu B\right\}$\\

$19$&${\rm Tr}\left\{\overline B \gamma^\mu\gamma^5 B\right\} {\rm Tr}\left\{u_\mu u_\nu u^\nu \right\}$\\
$20$&$i {\rm Tr}\left\{\overline B \gamma^\mu \left[\left[\left[D_\mu,u_\nu\right],u^\nu\right], B\right]\right\}$\\
$21$&$i {\rm Tr}\left\{\overline B \gamma^\mu \left\{\left[\left[D_\mu,u_\nu\right],u^\nu\right], B\right\}\right\}$\\
$22$&$ i\Big({\rm Tr}\left\{\overline B u_\nu\right\} \gamma^\mu{\rm Tr}\left\{\left[D_\mu,u^\nu\right]B\right\}-{\rm Tr}\left\{\overline B\left[D_\mu, u_\nu\right]\right\} \gamma^\mu{\rm Tr}\left\{u^\nu B\right\} \Big)$\\
$23$&$  \frac{i}{2 m_0}\Big( {\rm Tr}\left\{\overline B \sigma^{\mu \nu}\left[\left[D_\mu,u_\rho\right],\left[u_\nu,\left[D^\rho, B\right]\right]\right]\right\}+  {\rm Tr}\left\{\overline B   \sigma^{\mu \nu}\left[D_\rho,\left[u_\nu,\left[\left[D_\mu,u^\rho\right], B\right]\right]\right]\right\}   \Big)            $\\
$24$&$  \frac{i}{2 m_0}\Big( {\rm Tr}\left\{\overline B \sigma^{\mu \nu}\left[\left[D_\mu,u_\rho\right],\left\{u_\nu,\left[D^\rho, B\right]\right\}\right]\right\}+  {\rm Tr}\left\{\overline B   \sigma^{\mu \nu}\left[D_\rho,\left\{u_\nu,\left[\left[D_\mu,u^\rho\right], B\right]\right\}\right]\right\}   \Big)            $\\
$25$&$  \frac{i}{2 m_0}\Big( {\rm Tr}\left\{\overline B \sigma^{\mu \nu}\left\{\left[D_\mu,u_\rho\right],\left\{u_\nu,\left[D^\rho, B\right]\right\}\right\}\right\}+  {\rm Tr}\left\{\overline B   \sigma^{\mu \nu}\left[D_\rho,\left\{u_\nu,\left\{\left[D_\mu,u^\rho\right], B\right\}\right\}\right]\right\}   \Big)            $\\
$26$&$  \frac{i}{ m_0}\Big( {\rm Tr}\left\{\overline B \sigma^{\mu \nu} \left[D_\rho, B\right]\right\}{\rm Tr}\left\{\left[D_\mu,u^\rho\right] u_\nu\right\}+\frac{1}{2}{\rm Tr}\left\{\overline B \sigma^{\mu\nu}  B\right\}{\rm Tr}\left\{\left[D_\rho,\left[D_\mu,u^\rho\right]\right] u_\nu\right\}$\\&\quad$+\frac{1}{2}{\rm Tr}\left\{\overline B \sigma^{\mu\nu}  B\right\}{\rm Tr}\left\{\left[D_\mu,u_\rho\right] \left[D^\rho,u_\nu\right]\right\}   \Big)       $\\
$27$&$  \frac{i}{2 m_0^2} \Big(   {\rm Tr}\left\{\overline B \gamma^\mu\left[ \left[\left[D_\mu,u_\nu\right],u_\rho\right],\left[D^\nu,\left[D^\rho, B\right]\right]\right]\right\}$\\&\quad$+
 {\rm Tr}\left\{\overline B \gamma^\mu\left[D_\rho,\left[D_\nu,\left[ \left[\left[D_\mu,u^\nu\right],u^\rho\right] , B\right]\right]\right]\right\}     \Big)    $\\
$28$&$  \frac{i}{2 m_0^2} \Big(   {\rm Tr}\left\{\overline B \gamma^\mu\left\{ \left[\left[D_\mu,u_\nu\right],u_\rho\right],\left[D^\nu,\left[D^\rho, B\right]\right]\right\}\right\}$\\&\quad$+
 {\rm Tr}\left\{\overline B \gamma^\mu\left[D_\rho,\left[D_\nu,\left\{ \left[\left[D_\mu,u^\nu\right],u^\rho\right] , B\right\}\right]\right]\right\}     \Big)    $\\
$29$&$  \frac{i}{ m_0^2} \Big( {\rm Tr}\left\{\overline B u_\rho\right\} \gamma^\mu {\rm Tr}\left\{\left[D_\mu,u_\nu\right]\left[D^\nu,\left[D^\rho, B\right]\right]\right\}- {\rm Tr}\left\{\overline B\left[D_\mu,u_\nu\right]\right\} \gamma^\mu {\rm Tr}\left\{u_\rho\left[D^\rho, \left[D^\nu,B\right]\right]\right\}$ \\&\quad$-{\rm Tr}\left\{\overline B\left[D_\nu,\left[D_\mu,u^\nu\right]\right]\right\} \gamma^\mu {\rm Tr}\left\{u_\rho \left[D^\rho,B\right]\right\}-{\rm Tr}\left\{\overline B\left[D_\mu,u_\nu\right]\right\} \gamma^\mu {\rm Tr}\left\{\left[D^\nu,u_\rho\right]\left[D^\rho, B\right] \right\}$\\&\quad$-{\rm Tr}\left\{\overline B\left[D_\rho,\left[D_\mu,u_\nu\right]\right]\right\} \gamma^\mu {\rm Tr}\left\{u^\rho \left[D^\nu,B\right]\right\}-{\rm Tr}\left\{\overline B\left[D_\mu,u_\nu\right]\right\} \gamma^\mu {\rm Tr}\left\{\left[D_\rho,u^\rho\right]\left[D^\nu, B\right] \right\}$\\&\quad$ -{\rm Tr}\left\{\overline B\left[D_\rho,\left[D_\nu,\left[D_\mu,u^\nu\right]\right]\right]\right\} \gamma^\mu {\rm Tr}\left\{u^\rho B\right\}-{\rm Tr}\left\{\overline B\left[D_\rho,\left[D_\mu,u_\nu\right]\right]\right\} \gamma^\mu {\rm Tr}\left\{\left[D^\nu,u^\rho\right] B \right\}$\\&\quad$ -{\rm Tr}\left\{\overline B\left[D_\nu,\left[D_\mu,u^\nu\right]\right]\right\} \gamma^\mu {\rm Tr}\left\{\left[D_\rho,u^\rho\right] B\right\}-{\rm Tr}\left\{\overline B\left[D_\mu,u_\nu\right]\right\} \gamma^\mu {\rm Tr}\left\{\left[D_\rho,\left[D^\nu,u^\rho\right]\right] B \right\}    \Big)    $\\

$30$&$  \frac{1}{2 m_0} \Big({\rm Tr}\left\{\overline B \sigma^{\mu\nu}\gamma^5 \left[u_\mu,\left\{u_\nu,\left[u_\rho,\left[D^\rho, B\right]\right]\right\}\right]\right\}+{\rm Tr}\left\{\overline B\sigma^{\mu\nu}\gamma^5 \left[D_\rho,\left[u^\rho,\left\{u_\nu,\left[u_\mu, B\right]\right\}\right]\right]\right\}\Big)    $\\&
\\ \hline

\end{tabular}\end{center} \end{table*} 

\begin{table*}\begin{center}\begin{tabular}{|r|l|}

 \hline

&\\

$31$&$  \frac{1}{2 m_0} \Big({\rm Tr}\left\{\overline B \sigma^{\mu\nu}\gamma^5 \left[u_\mu,\left\{u_\nu,\left\{u_\rho,\left[D^\rho, B\right]\right\}\right\}\right]\right\}+{\rm Tr}\left\{\overline B\sigma^{\mu\nu}\gamma^5 \left[D_\rho,\left\{u^\rho,\left\{u_\nu,\left[u_\mu, B\right]\right\}\right\}\right]\right\}\Big)    $\\
$32$&$  \frac{1}{2 m_0} \Big( {\rm Tr}\left\{\overline B\sigma^{\mu\nu}\gamma^5\left[ \left[u_\mu,\left[u_\nu,u_\rho\right]\right],\left[D^\rho, B\right]\right]\right\}+ {\rm Tr}\left\{\overline B\sigma^{\mu\nu}\gamma^5 \left[D_\rho,\left[ \left[u_\mu,\left[u_\nu,u^\rho\right]\right], B\right]\right]\right\}  \Big)    $\\
$33$&$  \frac{1}{2 m_0} \Big( {\rm Tr}\left\{\overline B\sigma^{\mu\nu}\gamma^5\left\{ \left[u_\mu,\left[u_\nu,u_\rho\right]\right],\left[D^\rho, B\right]\right\}\right\}+ {\rm Tr}\left\{\overline B\sigma^{\mu\nu}\gamma^5 \left[D_\rho,\left\{ \left[u_\mu,\left[u_\nu,u^\rho\right]\right], B\right\}\right]\right\}  \Big)    $\\

$34$&$  \frac{1}{ m_0} \Big( {\rm Tr}\left\{\overline B u_\mu \right\}\sigma^{\mu\nu}\gamma^5 {\rm Tr}\left\{\left\{u_\nu,u_\rho\right\}\left[D^\rho, B\right]\right\}+ {\rm Tr}\left\{\overline B\left\{u_\nu,u_\rho\right\}\right\} \sigma^{\mu\nu}\gamma^5 {\rm Tr}\left\{u_\mu \left[D^\rho,B\right]\right\}$\\&\quad$+{\rm Tr}\left\{\overline B\left\{\left[D_\rho,u_\nu\right],u^\rho\right\}\right\}\sigma^{\mu\nu}\gamma^5 {\rm Tr}\left\{u_\mu B\right\}+{\rm Tr}\left\{\overline B\left\{u_\nu,\left[D_\rho,u^\rho\right]\right\}\right\}\sigma^{\mu\nu}\gamma^5 {\rm Tr}\left\{u_\mu B\right\}$\\&\quad$+{\rm Tr}\left\{\overline B\left\{u_\nu,u_\rho\right\}\right\}\sigma^{\mu\nu}\gamma^5 {\rm Tr}\left\{\left[D^\rho,u_\mu\right] B\right\}  \Big)    $\\

$35$&$  \frac{1}{ m_0} \Big( {\rm Tr}\left\{\overline B u_\rho\right\}\sigma^{\mu\nu}\gamma^5 {\rm Tr}\left\{\left[u_\mu,u_\nu\right]\left[D^\rho, B\right]\right\}- {\rm Tr}\left\{\overline B\left[u_\mu,u_\nu\right]\right\} \sigma^{\mu\nu}\gamma^5 {\rm Tr}\left\{u_\rho \left[D^\rho,B\right]\right\}$\\&\quad$-{\rm Tr}\left\{\overline B\left[\left[D_\rho,u_\mu\right],u_\nu\right]\right\}\sigma^{\mu\nu}\gamma^5 {\rm Tr}\left\{u^\rho B\right\}-{\rm Tr}\left\{\overline B\left[u_\mu,\left[D_\rho,u_\nu\right]\right]\right\}\sigma^{\mu\nu}\gamma^5 {\rm Tr}\left\{u^\rho B\right\}$\\&\quad$-{\rm Tr}\left\{\overline B\left[u_\mu,u_\nu\right]\right\}\sigma^{\mu\nu}\gamma^5 {\rm Tr}\left\{\left[D_\rho,u^\rho\right] B\right\}  \Big)    $\\

$36$&$  \frac{1}{2 m_0^2} \Big({\rm Tr}\left\{\overline B \gamma^\mu \gamma^5 \left[u_\mu,\left[u_\nu,\left[u_\rho,\left[D^\nu,\left[D^\rho, B\right]\right]\right]\right]\right]\right\}$\\&\quad$+{\rm Tr}\left\{\overline B \gamma^\mu \gamma^5\left[D_\rho,\left[D_\nu,\left[u^\rho,\left[u^\nu,\left[u_\mu, B\right]\right]\right]\right]\right]\right\} \Big)    $\\

$37$&$  \frac{1}{2 m_0^2} \Big({\rm Tr}\left\{\overline B \gamma^\mu \gamma^5 \left[u_\mu,\left\{u_\nu,\left[u_\rho,\left[D^\nu,\left[D^\rho, B\right]\right]\right]\right\}\right]\right\}$\\&\quad$+{\rm Tr}\left\{\overline B \gamma^\mu \gamma^5\left[D_\rho,\left[D_\nu,\left[u^\rho,\left\{u^\nu,\left[u_\mu, B\right]\right\}\right]\right]\right]\right\} \Big)    $\\

$38$&$  \frac{1}{2 m_0^2} \Big({\rm Tr}\left\{\overline B \gamma^\mu \gamma^5 \left\{u_\mu,\left[u_\nu,\left\{u_\rho,\left[D^\nu,\left[D^\rho, B\right]\right]\right\}\right]\right\}\right\}$\\&\quad$+{\rm Tr}\left\{\overline B \gamma^\mu \gamma^5\left[D_\rho,\left[D_\nu,\left\{u^\rho,\left[u^\nu,\left\{u_\mu, B\right\}\right]\right\}\right]\right]\right\} \Big)    $\\

$39$&$  \frac{1}{2 m_0^2} \Big({\rm Tr}\left\{\overline B \gamma^\mu \gamma^5 \left\{u_\mu,\left\{u_\nu,\left\{u_\rho,\left[D^\nu,\left[D^\rho, B\right]\right]\right\}\right\}\right\}\right\}$\\&\quad$+{\rm Tr}\left\{\overline B \gamma^\mu \gamma^5\left[D_\rho,\left[D_\nu,\left\{u^\rho,\left\{u^\nu,\left\{u_\mu, B\right\}\right\}\right\}\right]\right]\right\} \Big)    $\\

$40$&$  \frac{1}{ m_0^2} \Big( {\rm Tr}\left\{\overline B \gamma^\mu \gamma^5  \left[D_{\nu },\left[D_\rho,B\right]\right]\right\} {\rm Tr}\left\{u_\mu \left\{u^\nu,u^\rho\right\}\right\}$\\&\quad$+{\rm Tr}\left\{\overline B \gamma^\mu \gamma^5\left[D_\nu, B\right]\right\} {\rm Tr}\left\{\left[D_\rho,u_\mu\right] \left\{u^\nu,u^\rho\right\}\right\}$\\&\quad$+{\rm Tr}\left\{\overline B \gamma^\mu \gamma^5 \left[D_\nu, B\right]\right\} {\rm Tr}\left\{u_\mu \left\{\left[D_\rho,u^\nu\right],u^\rho\right\}\right\}$\\&\quad$+{\rm Tr}\left\{\overline B \gamma^\mu \gamma^5 \left[D_\nu, B\right]\right\} {\rm Tr}\left\{u_\mu \left\{u^\nu,\left[D_\rho,u^\rho\right]\right\}\right\} \Big)$\\
$41$&${\rm Tr}\left\{\overline B \gamma^\mu \gamma^5 \left[u_\mu,\left[\chi^+, B\right]\right]\right\}+{\rm Tr}\left\{\overline B\gamma^\mu \gamma^5 \left[\chi^+,\left[u_\mu, B\right]\right]\right\} $\\
$42$&${\rm Tr}\left\{\overline B \gamma^\mu \gamma^5 \left[u_\mu,\left\{\chi^+, B\right\}\right]\right\}+{\rm Tr}\left\{\overline B\gamma^\mu \gamma^5 \left\{\chi^+,\left[u_\mu, B\right]\right\}\right\} $\\
$43$&${\rm Tr}\left\{\overline B \gamma^\mu \gamma^5 \left\{u_\mu,\left[\chi^+, B\right]\right\}\right\}+{\rm Tr}\left\{\overline B\gamma^\mu \gamma^5 \left[\chi^+,\left\{u_\mu, B\right\}\right]\right\} $\\
$44$&${\rm Tr}\left\{\overline B \gamma^\mu \gamma^5 \left\{u_\mu,\left\{\chi^+, B\right\}\right\}\right\}+{\rm Tr}\left\{\overline B\gamma^\mu \gamma^5 \left\{\chi^+,\left\{u_\mu, B\right\}\right\}\right\} $\\ 

$45$&${\rm Tr}\left\{\overline B\gamma^\mu \gamma^5  \left[u_\mu,B\right]\right\} {\rm Tr}\left\{\chi^+\right\}$\\ 
$46$&${\rm Tr}\left\{\overline B\gamma^\mu \gamma^5  \left\{u_\mu,B\right\}\right\} {\rm Tr}\left\{\chi^+\right\}$\\

$47$&$ {\rm Tr}\left\{\overline B\gamma^\mu \gamma^5  B\right\} {\rm Tr}\left\{u_\mu \chi^+\right\}$\\

$48$&${\rm Tr}\left\{\overline B \gamma^\mu \left[\left[u_\mu,\chi^-\right], B\right]\right\}$\\
$49$&${\rm Tr}\left\{\overline B \gamma^\mu \left\{\left[u_\mu,\chi^-\right], B\right\}\right\}$\\
$50$&$ {\rm Tr}\left\{\overline B u_\mu\right\} \gamma^\mu {\rm Tr}\left\{\chi^- B\right\}- {\rm Tr}\left\{\overline B \chi^-\right\} \gamma^\mu {\rm Tr}\left\{u_\mu B\right\}$\\
$51$&$  {\rm Tr}\left\{\overline B \gamma^\mu \left[\left[D^\nu,F^+_{\mu \nu}\right], B\right]\right\}$\\
$52$&$  {\rm Tr}\left\{\overline B \gamma^\mu \left\{\left[D^\nu,F^+_{\mu \nu}\right], B\right\}\right\}$\\
$53$&${\rm Tr}\left\{\overline B\varepsilon^{\mu\nu\rho\tau}\gamma_\tau \left[u_\mu,\left[F^+_{\nu \rho}, B\right]\right]\right\}+{\rm Tr}\left\{\overline B\varepsilon^{\mu\nu\rho\tau}\gamma_\tau \left[F^+_{\nu \rho},\left[u_\mu, B\right]\right]\right\}$\\

$54$&${\rm Tr}\left\{\overline B\varepsilon^{\mu\nu\rho\tau}\gamma_\tau \left[u_\mu,\left\{F^+_{\nu \rho}, B\right\}\right]\right\}+{\rm Tr}\left\{\overline B\varepsilon^{\mu\nu\rho\tau}\gamma_\tau \left\{F^+_{\nu \rho},\left[u_\mu, B\right]\right\}\right\}$\\

$55$&${\rm Tr}\left\{\overline B\varepsilon^{\mu\nu\rho\tau}\gamma_\tau \left\{u_\mu,\left[F^+_{\nu \rho}, B\right]\right\}\right\}+{\rm Tr}\left\{\overline B\varepsilon^{\mu\nu\rho\tau}\gamma_\tau \left[F^+_{\nu \rho},\left\{u_\mu, B\right\}\right])\right\}$\\

$56$&${\rm Tr}\left\{\overline B\varepsilon^{\mu\nu\rho\tau}\gamma_\tau \left\{u_\mu,\left\{F^+_{\nu \rho}, B\right\}\right\}\right\}+{\rm Tr}\left\{\overline B\varepsilon^{\mu\nu\rho\tau}\gamma_\tau \left\{F^+_{\nu \rho},\left\{u_\mu, B\right\}\right\}\right\}$\\ 

$57$&${\rm Tr}\left\{\overline B\varepsilon^{\mu\nu\rho\tau}\gamma_\tau  B\right\} {\rm Tr}\left\{u_\mu F^+_{\nu \rho}\right\}$\\&\\
 \hline

\end{tabular}\end{center} \end{table*} 

\begin{table*}\begin{center}\begin{tabular}{|r|l|}

 \hline &\\

$58$&$i{\rm Tr}\left\{\overline B \gamma^\mu \gamma^5 \left[\left[u^\nu,F^+_{\mu\nu}\right], B\right]\right\}$\\

$59$&$i{\rm Tr}\left\{\overline B \gamma^\mu \gamma^5 \left\{\left[u^\nu,F^+_{\mu\nu}\right], B\right\}\right\}$\\

$60$&$i\Big({\rm Tr}\left\{\overline B u^\nu\right\}\gamma^\mu \gamma^5  {\rm Tr}\left\{F^+_{\mu\nu} B\right\}- {\rm Tr}\left\{\overline B F^+_{\mu\nu} \right\}\gamma^\mu \gamma^5  {\rm Tr}\left\{u^\nu B\right\}\Big)$\\
$61$&$\frac{i}{2m_0}\Big({\rm Tr}\left\{\overline B \sigma^{\mu\nu}\gamma^5 \left[\left[u_\mu,F^+_{\nu \rho}\right],\left[D^\rho, B\right]\right]\right\}+ {\rm Tr}\left\{\overline B \sigma^{\mu\nu}\gamma^5 \left[D^\rho,\left[\left[u_\mu,F^+_{\nu\rho}\right], B\right]\right]\right\}\Big)$\\

$62$&$\frac{i}{2m_0}\Big({\rm Tr}\left\{\overline B \sigma^{\mu\nu}\gamma^5 \left\{\left[u_\mu,F^+_{\nu \rho}\right],\left[D^\rho, B\right]\right\}\right\}+ {\rm Tr}\left\{\overline B \sigma^{\mu\nu}\gamma^5 \left[D^\rho,\left\{\left[u_\mu,F^+_{\nu\rho}\right], B\right\}\right]\right\}\Big)$\\

$63$&$\frac{i}{m_0}\Big( {\rm Tr}\left\{\overline B u_\mu\right\} \sigma^{\mu\nu}\gamma^5 {\rm Tr}\left\{F^+_{\nu\rho} \left[D^\rho, B\right]\right\}-{\rm Tr}\left\{\overline B F^+_{\nu\rho}\right\} \sigma^{\mu\nu}\gamma^5 {\rm Tr}\left\{u_\mu \left[D^\rho, B\right]\right\}$\\&\quad$- {\rm Tr}\left\{\overline B\left[D^\rho,F^+_{\nu\rho}\right]  \right\} \sigma^{\mu\nu}\gamma^5 {\rm Tr}\left\{u_\mu B\right\}- {\rm Tr}\left\{\overline BF^+_{\nu\rho}\right\} \sigma^{\mu\nu}\gamma^5 {\rm Tr}\left\{\left[D^\rho,u_\mu\right] B\right\}\Big)$\\
$64$&$  {\rm Tr}\left\{\overline B \gamma^\mu \gamma^5\left[\left[D^\nu,F^-_{\mu \nu}\right], B\right]\right\}$\\
$65$&$  {\rm Tr}\left\{\overline B \gamma^\mu\gamma^5 \left\{\left[D^\nu,F^-_{\mu \nu}\right], B\right\}\right\}$\\

$66$&${\rm Tr}\left\{\overline B\varepsilon^{\mu\nu\rho\tau}\gamma_\tau\gamma^5 \left[u_\mu,\left[F^-_{\nu \rho}, B\right]\right]\right\}+{\rm Tr}\left\{\overline B\varepsilon^{\mu\nu\rho\tau}\gamma_\tau\gamma^5 \left[F^-_{\nu \rho},\left[u_\mu, B\right]\right]\right\}$\\

$67$&${\rm Tr}\left\{\overline B\varepsilon^{\mu\nu\rho\tau}\gamma_\tau\gamma^5 \left[u_\mu,\left\{F^-_{\nu \rho}, B\right\}\right]\right\}+{\rm Tr}\left\{\overline B\varepsilon^{\mu\nu\rho\tau}\gamma_\tau\gamma^5 \left\{F^-_{\nu \rho},\left[u_\mu, B\right]\right\}\right\}$\\

$68$&${\rm Tr}\left\{\overline B\varepsilon^{\mu\nu\rho\tau}\gamma_\tau \gamma^5\left\{u_\mu,\left[F^-_{\nu \rho}, B\right]\right\}\right\}+{\rm Tr}\left\{\overline B\varepsilon^{\mu\nu\rho\tau}\gamma_\tau\gamma^5 \left[F^-_{\nu \rho},\left\{u_\mu, B\right\}\right])\right\}$\\

$69$&${\rm Tr}\left\{\overline B\varepsilon^{\mu\nu\rho\tau}\gamma_\tau\gamma^5 \left\{u_\mu,\left\{F^-_{\nu \rho}, B\right\}\right\}\right\}+{\rm Tr}\left\{\overline B\varepsilon^{\mu\nu\rho\tau}\gamma_\tau\gamma^5 \left\{F^-_{\nu \rho},\left\{u_\mu, B\right\}\right\}\right\}$\\ 

$70$&${\rm Tr}\left\{\overline B\varepsilon^{\mu\nu\rho\tau}\gamma_\tau\gamma^5  B\right\} {\rm Tr}\left\{u_\mu F^-_{\nu \rho}\right\}$\\

$71$&$i{\rm Tr}\left\{\overline B \gamma^\mu  \left[\left[u^\nu,F^-_{\mu\nu}\right], B\right]\right\}$\\

$72$&$i{\rm Tr}\left\{\overline B \gamma^\mu  \left\{\left[u^\nu,F^-_{\mu\nu}\right], B\right\}\right\}$\\

$73$&$i\Big({\rm Tr}\left\{\overline B u^\nu\right\}\gamma^\mu   {\rm Tr}\left\{F^-_{\mu\nu} B\right\}- {\rm Tr}\left\{\overline B F^-_{\mu\nu} \right\}\gamma^\mu   {\rm Tr}\left\{u^\nu B\right\}\Big)$\\

$74$&$\frac{i}{2m_0}\Big( {\rm Tr}\left\{\overline B \sigma^{\mu\nu} \left[u_\mu,\left[F^-_{\nu\rho},\left[D^\rho, B\right]\right]\right]\right\}+{\rm Tr}\left\{\overline B\sigma^{\mu\nu}\left[D^\rho,\left[F^-_{\nu\rho},\left[u_\mu, B\right]\right]\right]\right\}\Big)$\\

$75$&$\frac{i}{2m_0}\Big( {\rm Tr}\left\{\overline B \sigma^{\mu\nu} \left[u_\mu,\left\{F^-_{\nu\rho},\left[D^\rho, B\right]\right\}\right]\right\}+{\rm Tr}\left\{\overline B\sigma^{\mu\nu}\left[D^\rho,\left\{F^-_{\nu\rho},\left[u_\mu, B\right]\right\}\right]\right\}\Big)$\\

$76$&$\frac{i}{2m_0}\Big( {\rm Tr}\left\{\overline B \sigma^{\mu\nu} \left\{u_\mu,\left[F^-_{\nu\rho},\left[D^\rho, B\right]\right]\right\}\right\}+{\rm Tr}\left\{\overline B\sigma^{\mu\nu}\left[D^\rho,\left[F^-_{\nu\rho},\left\{u_\mu, B\right\}\right]\right]\right\}\Big)$\\

$77$&$\frac{i}{2m_0}\Big( {\rm Tr}\left\{\overline B \sigma^{\mu\nu} \left\{u_\mu,\left\{F^-_{\nu\rho},\left[D^\rho, B\right]\right\}\right\}\right\}+{\rm Tr}\left\{\overline B\sigma^{\mu\nu}\left[D^\rho,\left\{F^-_{\nu\rho},\left\{u_\mu, B\right\}\right\}\right]\right\}\Big)$\\

$78$&$\frac{i}{m_0}\Big( {\rm Tr}\left\{\overline B \sigma^{\mu\nu}  \left[D^\rho,B\right]\right\} {\rm Tr}\left\{u_\mu F^-_{\nu\rho}\right\}+\frac{1}{2}{\rm Tr}\left\{\overline B \sigma^{\mu\nu}  B\right\} {\rm Tr}\left\{\left[D^\rho,u_\mu\right]F^-_{\nu\rho}\right\}$\\&\quad$+\frac{1}{2}{\rm Tr}\left\{\overline B \sigma^{\mu\nu}  B\right\} {\rm Tr}\left\{u_\mu \left[D^\rho,F^-_{\nu\rho}\right]\right\}  \Big)$\\

&\\ \hline

\end{tabular}
\vspace{0.4cm}
\caption{Operators of ${\cal L}^{(3)}_{MB}\,$ For definitions, 
see Refs.\cite{Oller:2006yh,Fettes:2000gb}.}
\label{table:1}
\end{center} \end{table*}



\begin{thebibliography}{99} 

\bibitem{Oller:2006yh}
J.~A.~Oller, M.~Verbeni and J.~Prades,
arXiv:hep-ph/0608204.

\bibitem{Fettes:2000gb}
N.~Fettes, U.-G.~Mei{\ss}ner, M.~Moj\v zi\v s and S.~Steininger,
Annals Phys.\  {\bf 283} (2000) 273
[Erratum-ibid.\  {\bf 288} (2001) 249]
[arXiv:hep-ph/0001308].

\bibitem{Krause:1990xc}
A.~Krause,
Helv.\ Phys.\ Acta {\bf 63} (1990) 3.

\bibitem{FM4}
M.~Frink and U.-G.~Mei{\ss}ner, ``The chiral effective meson-baryon Lagrangian
at fourth order,'' {\em in preparation}.

\bibitem{Frink:2004ic}
M.~Frink and U.-G.~Mei{\ss}ner,
JHEP {\bf 0407} (2004) 028
[arXiv:hep-lat/0404018].


\end{thebibliography}
\end{document}